\documentclass[%
reprint,
superscriptaddress,
amsmath,amssymb,
prl,
]{revtex4-2}
\usepackage{color}
\usepackage{float}
\usepackage[pdftex]{graphicx}
\usepackage{graphicx}% Include figure files
\usepackage{dcolumn}% Align table columns on decimal point
\usepackage{bm}% bold math
\usepackage{siunitx}
\usepackage{comment}
\usepackage{fixltx2e}
\usepackage{soul}
\usepackage{braket}

\usepackage{hyperref}% add hypertext capabilities
\hypersetup{
    bookmarks=true,         % show bookmarks bar?
    unicode=false,          % non-Latin characters in Acrobat’s bookmarks
    pdftoolbar=true,        % show Acrobat’s toolbar?
    pdfmenubar=true,        % show Acrobat’s menu?
    pdffitwindow=false,     % window fit to page when opened
    pdftitle={Probing charge noise in few electron CMOS quantum dots},    % title
    pdfauthor={Dr. Matias Urdampilleta},     % author
    pdfsubject={},   % subject of the document
    pdfcreator={Dr. Matias Urdampilleta},   % creator of the document
    pdfproducer={},  % producer of the document
    pdfkeywords={,} {} {}, % list of keywords
    pdfnewwindow=true,      % links in new window
    colorlinks=false,       % false: boxed links; true: colored links
    linkcolor=red,          % color of internal links
    citecolor=green,        % color of links to bibliography
    filecolor=magenta,      % color of file links
    urlcolor=cyan           % color of external links
}
\usepackage{filecontents}
\usepackage{comment}

\DeclareUnicodeCharacter{2009}{\,} 
\DeclareSIUnit{\belmilliwatt}{Bm}
\DeclareSIUnit{\dBm}{\deci\belmilliwatt}
\newcommand{\doublecolumn}{176mm}
\begin{document}
\bibliographystyle{apsrev4-1}

\preprint{APS}
%\title{Charge detection in an array of CMOS quantum dots}% Force line breaks with \\
\title{Probing low-frequency charge noise in few-electron CMOS quantum dots}% Force line breaks with \\
 \author{Cameron Spence}
\email{cameron.spence@neel.cnrs.fr}
\affiliation{Univ. Grenoble Alpes, CNRS, Grenoble INP, Institut N\'eel, 38402 Grenoble, France}

\author{Bruna Cardoso Paz}
\affiliation{Univ. Grenoble Alpes, CNRS, Grenoble INP, Institut N\'eel, 38402 Grenoble, France}

\author{Vincent Michal}
\affiliation{Univ. Grenoble Alpes, CEA, IRIG, 38000 Grenoble, France}

\author{Emmanuel Chanrion}
\affiliation{Univ. Grenoble Alpes, CNRS, Grenoble INP, Institut N\'eel, 38402 Grenoble, France}

\author{David J. Niegemann}
\affiliation{Univ. Grenoble Alpes, CNRS, Grenoble INP, Institut N\'eel, 38402 Grenoble, France}

\author{Baptiste Jadot}
\affiliation{Univ. Grenoble Alpes, CEA, Leti, F-38000 Grenoble, France}

\author{Pierre-Andr\'e Mortemousque}
\affiliation{Univ. Grenoble Alpes, CEA, Leti, F-38000 Grenoble, France}

\author{Bernhard Klemt}
\affiliation{Univ. Grenoble Alpes, CNRS, Grenoble INP, Institut N\'eel, 38402 Grenoble, France}

\author{Vivien Thiney}
\affiliation{Univ. Grenoble Alpes, CNRS, Grenoble INP, Institut N\'eel, 38402 Grenoble, France}

\author{Benoit Bertrand}
\affiliation{Univ. Grenoble Alpes, CEA, Leti, F-38000 Grenoble, France}

\author{Louis Hutin}
\affiliation{Univ. Grenoble Alpes, CEA, Leti, F-38000 Grenoble, France}

\author{Christopher B{\"a}uerle}
\affiliation{Univ. Grenoble Alpes, CNRS, Grenoble INP, Institut N\'eel, 38402 Grenoble, France}

\author{Maud Vinet}
\affiliation{Univ. Grenoble Alpes, CEA, Leti, F-38000 Grenoble, France}

\author{Yann-Michel Niquet}
\affiliation{Univ. Grenoble Alpes, CEA, IRIG, 38000 Grenoble, France}

\author{Tristan Meunier}
\affiliation{Univ. Grenoble Alpes, CNRS, Grenoble INP, Institut N\'eel, 38402 Grenoble, France}

\author{Matias Urdampilleta	}
\email{matias.urdampilleta@neel.cnrs.fr}
\affiliation{Univ. Grenoble Alpes, CNRS, Grenoble INP, Institut N\'eel, 38402 Grenoble, France}

\date{\today}% It is always \today, today,
\begin{abstract}
      	Charge noise is one of the main sources of environmental decoherence for spin qubits in silicon, presenting a major obstacle in the path towards highly scalable and reproducible qubit fabrication.
	Here we demonstrate in-depth characterization of the charge noise environment experienced by a quantum dot in a CMOS-fabricated silicon nanowire.
	We probe the charge noise for different quantum dot configurations, finding that it is possible to tune the charge noise over two orders of magnitude, ranging from $\SI{1}{\micro eV^2/Hz}$ to $\SI{100}{\micro eV^2/Hz}$. In particular, we show that the top interface and the reservoirs are the main sources of charge noise and their effect can be mitigated by controlling the quantum dot extension.
	Additionally, we demonstrate a novel method for the measurement of the charge noise experienced by a quantum dot in the few electron regime.
	We measure a comparatively high charge noise value of $\SI{40}{\micro eV^2/Hz}$ at the first electron, and demonstrate that the charge noise is highly dependent on the electron occupancy of the quantum dot.
\end{abstract}

\maketitle

%%%%%%%%%%%%%%%%%%%%%%%%%%%%%%%%%%%%%%%%%%%%%%%%%%%%%%%%%%%%%%%%%%%%%%%%
\section{Introduction}

Spin qubits in silicon are a strong potential candidate for the eventual qubit of a scalable quantum computer \cite{Kane1998}.
State-of-the-art research has demonstrated long spin coherence times \cite{muhonen2014}, single qubit gate fidelity above 99.9$\%$ \cite{yoneda2018quantum}, and two-gate fidelity over 99$\%$ \cite{noiri2022fast} in silicon quantum dots.
However, the coherence time of silicon qubits can be limited by combined natural or induced spin-orbit coupling and low frequency $1/f$ charge noise \cite{yoneda2018quantum,Culcer2009}.
This charge noise, induced by local electric field fluctuations, is a characteristic of the material and architecture used to form the qubit.
In semiconductor devices, it is believed to be dominated by bistable charge traps, in particular dangling bonds at the semiconductor/oxide interface or dopants in the active regions \cite{connors2019}.
Quantum dots in silicon are tightly confined by material interfaces and therefore are sensitive to charge noise \cite{chanrion2020charge}.

Spin measurement and manipulation techniques have been shown to be compatible with industrially-produced 300 mm technology silicon nanowire devices \cite{Urdampilleta2019, ciriano2021spin, DzurakReflecto, maurand2016, crippa2019}. 
However, these device geometries and fabrication processes are different from heterostructure-based devices, and constrain quantum dots with additional interfaces, providing a potential source of increased charge noise.
This is particularly relevant in the few electron regime where the quantum dots are more sensitive to Coulomb disorder, and direct characterization of charge noise can be challenging. %Add ref
Characterizing the charge noise is important to direct fabrication and develop methods to mitigate its effect on qubits \cite{kestner2013noise}.

In this letter, we present the extended characterization of CMOS quantum dots fabricated on a foundry-compatible 300 mm wafer.
In particular, we first show how we can probe the different sources of noise by modifying the electronic wave function.
We find that the level of charge noise depends drastically on shape and position of the quantum dot within the channel.
In the second part, we introduce a method to extract charge noise and its power spectral density at the single electron level.
This last result is of particular importance to understand the role of charge noise in few-electron quantum dots, which is the relevant regime to operate as spin qubits.
It enables simple measurements of the fluctuation in chemical potential, which is traditionally extracted via more complex experiments \cite{yoneda2018quantum, struck2020, connors2021charge}.

\section{Device design and operation}

\begin{figure}%
\includegraphics[width=\columnwidth]{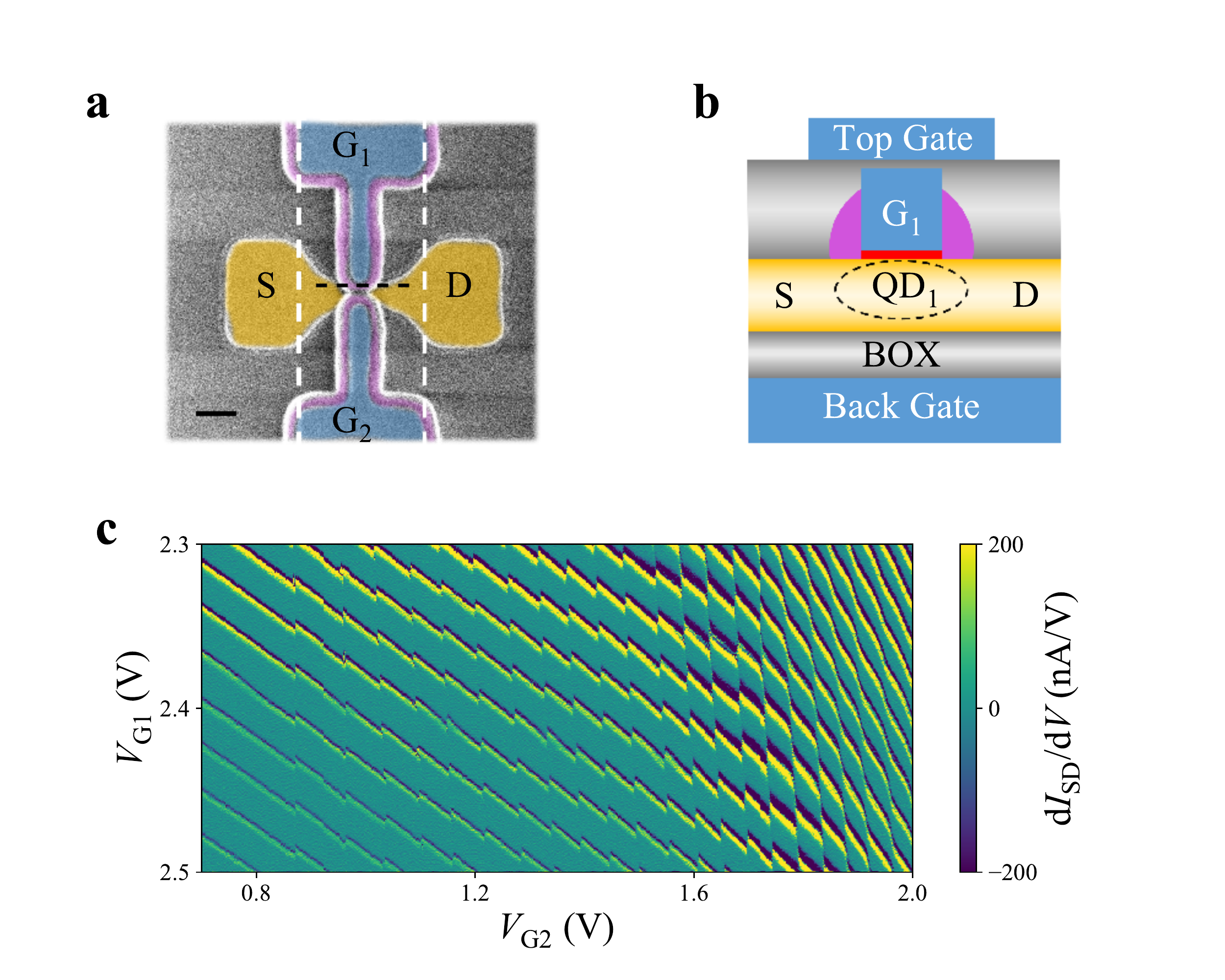}
\caption{
(a) False-colour SEM image of an identical CMOS nanowire to the one used to form quantum dots.
The nanowire channel (yellow) is enveloped by electrostatic gates (blue), which are separated by nitride spacers (pink).
The spacers screen the channel from the high density doping used to define the source and drain. The scale bar is \SI{100}{nm}.
(b) Cross sectional schematic (not to scale) of the nanowire along the black dashed line in (a).
(c) Two-dimensional stability diagram demonstrating charge sensing down to the first electron under G2.
The sharp cuts in the conductance peaks indicate the capacitive shift induced by the addition of an electron in the dot under G2.
}
\label{fig:device}%	
\end{figure}

  \begin{figure}%
  \includegraphics[width=\columnwidth]{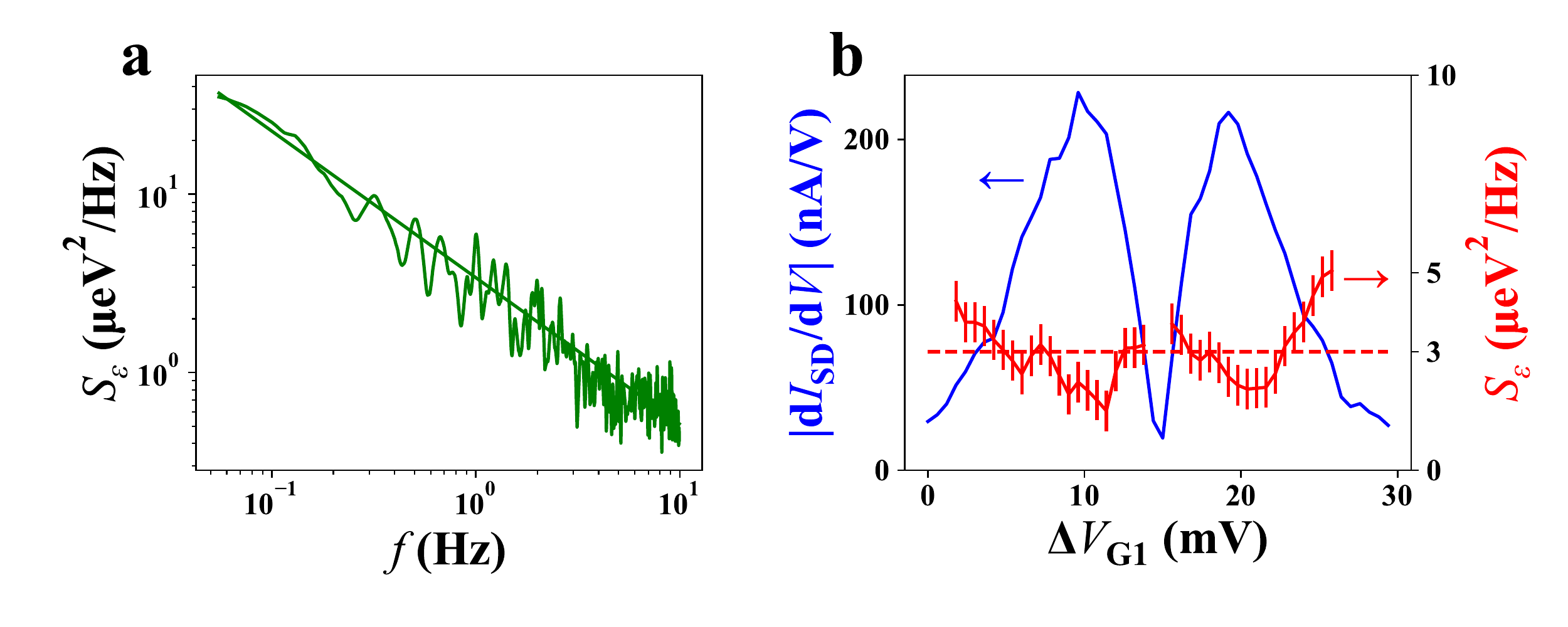}
  \caption{
(a) $1/f$-like power density spectrum acquired via Fourier transform of a current time trace, measured on the slope of a coulomb peak.
(b) Method for extracting the average charge noise value across a coulomb peak.
Current time traces are measured for varying values of $V_{G1}$ across a coulomb peak.
The shape of the coulomb peak is extracted from the average current value for each noise trace.
The absolute conductance of the peak (blue) gives the sensitivity of the detector dot for each value of gate voltage.
This is used to renormalize the PSD measured at each gate voltage value.
Spectra where the conductance is below $\SI{50}{nS}$ are discarded due to low sensitivity and over-renormalization leading to skewed results.
The magnitude of the PSD at $\SI{1}{Hz}$ is extracted from a fit of the noise spectrum at acquisition points spanning the coulomb peak (red).
The average charge noise value for this measurement is indicated by the dashed line at $\SI{3}{\micro eV^2 / Hz}$.
}
  \label{fig:1/f}%	
  \end{figure}

The device under study is a split-gate nanowire CMOS design fabricated on a foundry-compatible 300 mm wafer, identical to that shown in Fig. \ref{fig:device}(a).
The device design involves a semiconducting silicon channel which is covered by two parallel electrostatic plunger gates, marked $G_1$ and $G_2$. 
The channel has a width of $\SI{100}{\nano\metre}$, a gate length of $\SI{60}{\nano\metre}$, and a vertical split of $\SI{60}{\nano\metre}$.
The plunger gates are insulated from the channel by a resistive layer of native oxide and titanium nitride with a thickness of a few nanometres, marked in red in Fig. \ref{fig:device}(b).
The source and drain reservoirs are formed by high density implantation of dopants in the areas of the channel not covered by the silicon nitride spacers (indicated in pink).
A homogeneous back gate is formed by the silicon bulk below the buried oxide.
At low temperature, charge freeze-out necessitates using an LED to generate carriers in the bulk to polarize the back gate.
Tests in the same configuration with and without the LED indicate that photoexcitation does not significantly contribute to the charge noise.
In addition, a metal top gate is deposited above the plunger gates for finer control of the electrostatic environment of the channel.

The device has been measured in a $^3$He refrigerator, with a base temperature of $\SI{300}{\milli\kelvin}$.
At this temperature, a quantum dot (QD) is formed under each front gate, G$_1$ and G$_2$. 
Additionally, the spatial extension and coupling to reservoirs of the quantum dots can be further tuned by applying voltage on the top gate and on the back gate, see Fig. \ref{fig:device}(b).
The identification of the number of electrons in each dot is achieved by operating one of the two quantum dots as a charge sensor.
Fig. \ref{fig:device}(c) shows the stability diagram of $V_{G1}$, $V_{G2}$, which indicates the number of electrons in QD2, using QD1 as a charge sensor.

Charge noise measurements are performed by recording the current flowing through one of the quantum dots as a function of time.
A Fourier transform of the time trace gives the power spectral density (PSD) of the signal.
The current noise spectrum is then renormalized by $dI/dV_G$ and the lever arm $\alpha$ to account for the sensitivity of the detector at the measurement point.
We can then plot the charge noise $S_\epsilon$ as a function of frequency, see Fig.\ref{fig:1/f}(a).
Fig. \ref{fig:1/f}(b) shows that the extracted charge noise at $\SI{1}{Hz}$ does not vary much over a Coulomb peak.
The average value is close to $3~\si{\micro\electronvolt}^2$/Hz at $\SI{360}{mK}$ which is a relatively low value for silicon compared to other values obtained in literature \cite{connors2021charge,jock2018silicon,mi2018landau,kim2019low}.
We attribute the small fluctuations to a minor effect of the plunger gate on the electrostatic environment of the quantum dot.

%%%%%%%%%%%%%%%%%%%%%%%%%%%%%%%%%%%%%%%%%%%%%%%%%%%%%%%%%%%%%%%%%%%%%%%%
\section{Probing the noise sources in the many-electron regime}
One of the main challenges when assessing the noise environment is disentangling the different sources of noise \cite{you2015, kamioka2014}.
In a device where the accumulation interface is expected to be the dominating source of noise, it is important to make a distinction with other contributions such as noise originating from the core of the nanowire, reservoirs or other interfaces.
For this purpose, we can distort the quantum dot confinement potential in order to probe the spatial dependence of charge noise on a nanometre scale.

\begin{figure}%
\includegraphics[scale=0.39]{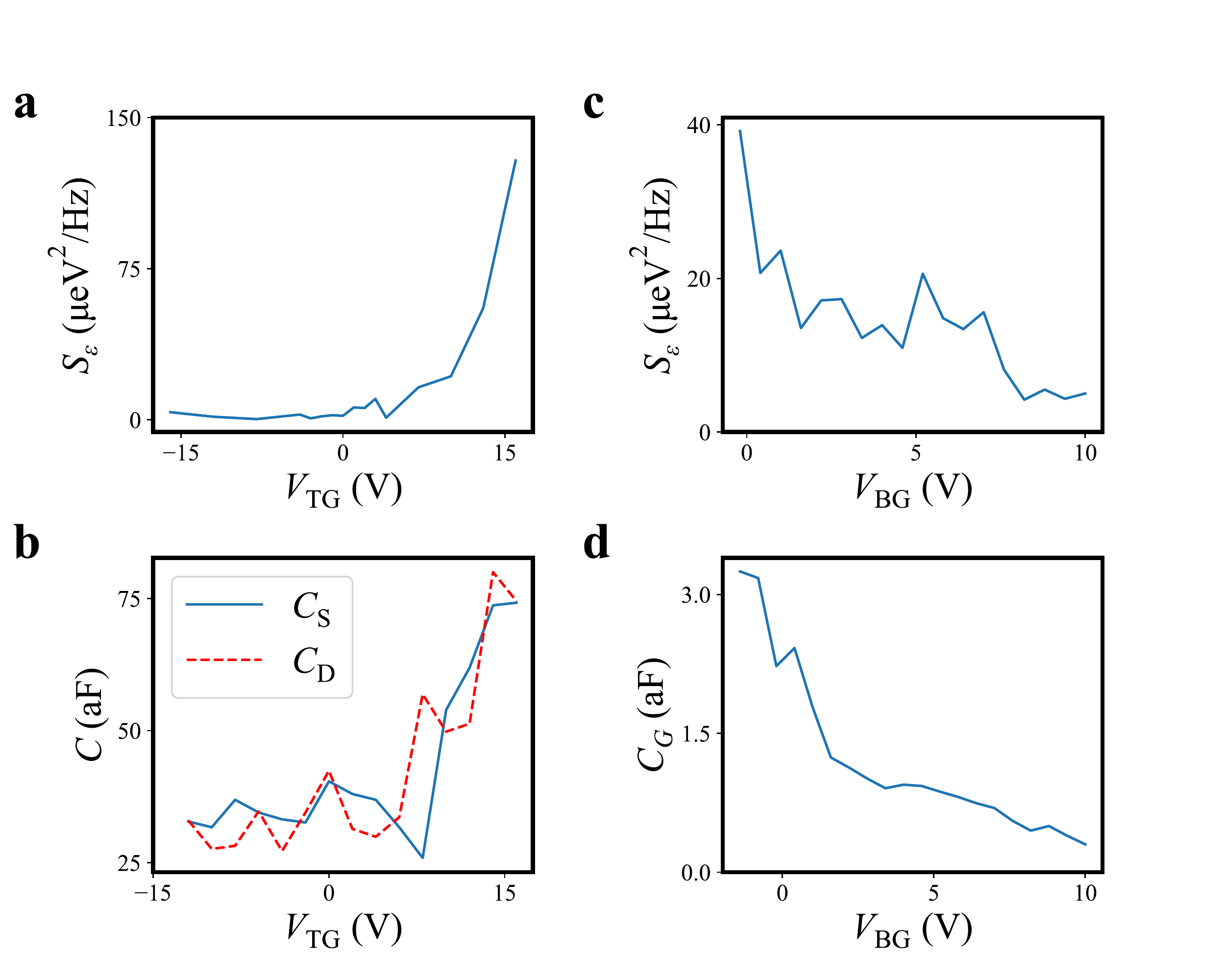}%
\caption{
(a) Power spectral density value at $\SI{1}{Hz}$ as a function of the top gate voltage.
The number of electrons in the dot is constant for all values of $V_{TG}$.
(b) Variation in the reservoir capacitances to the quantum dot as the top gate voltage is varied, measured simultaneously with the PSD in (a).
(c) Power spectral density at $\SI{1}{Hz}$ measured as a function of the back gate voltage.
The number of electrons in the dot remains fixed.
(d) Variation in the plunger gate capacitance as a function of the back gate voltage.
A lower capacitance indicates a weaker coupling to the plunger gate as the quantum dot moves vertically in the channel.
}
\label{fig:capa}%	
\end{figure}

We start by studying the effect of the back-gate which allows us to move the QD vertically within the channel.
By applying a positive voltage on the back gate, the QD which was originally confined against the top interface, now moves towards the back interface.
This is visible in Fig. \ref{fig:capa}(d), where the front gate to QD capacitance is plotted as a function of the back gate.
In parallel, we record the evolution of charge noise with back gate, see Fig. \ref{fig:capa}(c). 
We observe that charge noise decreases as the QD is less confined against the top interface.
This is in good agreement with literature which reports a large density of charge traps at the Si/SiO$_2$ gate interface \cite{kim2017annealing, freeman2016}.
This is also in qualitative agreement with the simulation of the effect of a single TLS above the interface on the chemical potential of the quantum dot as a function of back gate voltage (see supplementary material)\ref{supinf:simu}.

The reservoirs provide electrons to the quantum dot, but can also be a source of charge noise due to slow-varying fluctuators at the diffuse boundary.
To investigate the effect of the reservoirs, we first bias the back gate in a region of relatively weak charge noise ($V_{BG}$=\SI{10}{V}).
We then use the top gate to extend the QD wavefunction towards the reservoirs, whilst accounting for unwanted vertical extension by applying a compensating voltage on the plunger gate.
Fig. \ref{fig:capa}(a) and (b) present the evolution of the charge noise, and source and drain capacitance, respectively as a function of the top gate voltage.
It shows that as the capacitive coupling to reservoirs increases, the charge noise experienced by the QD likewise increases in magnitude.
Therefore, the reservoirs can also become the dominant source of noise either by directly modifying the chemical potential of the QD or by modifying the tunnel barrier transparency.
The main source of noise in the reservoirs is expected to be dopants that have diffused on the side of the tunnel junction, where the reservoirs stop being metallic.
These dopants can act as two-level fluctuators, and are expected to fluctuate at a rather high frequency due to their relatively large coupling to the electron reservoir.
% Add a passage about the spacers?

\begin{figure*}%
  \includegraphics[width=\doublecolumn]{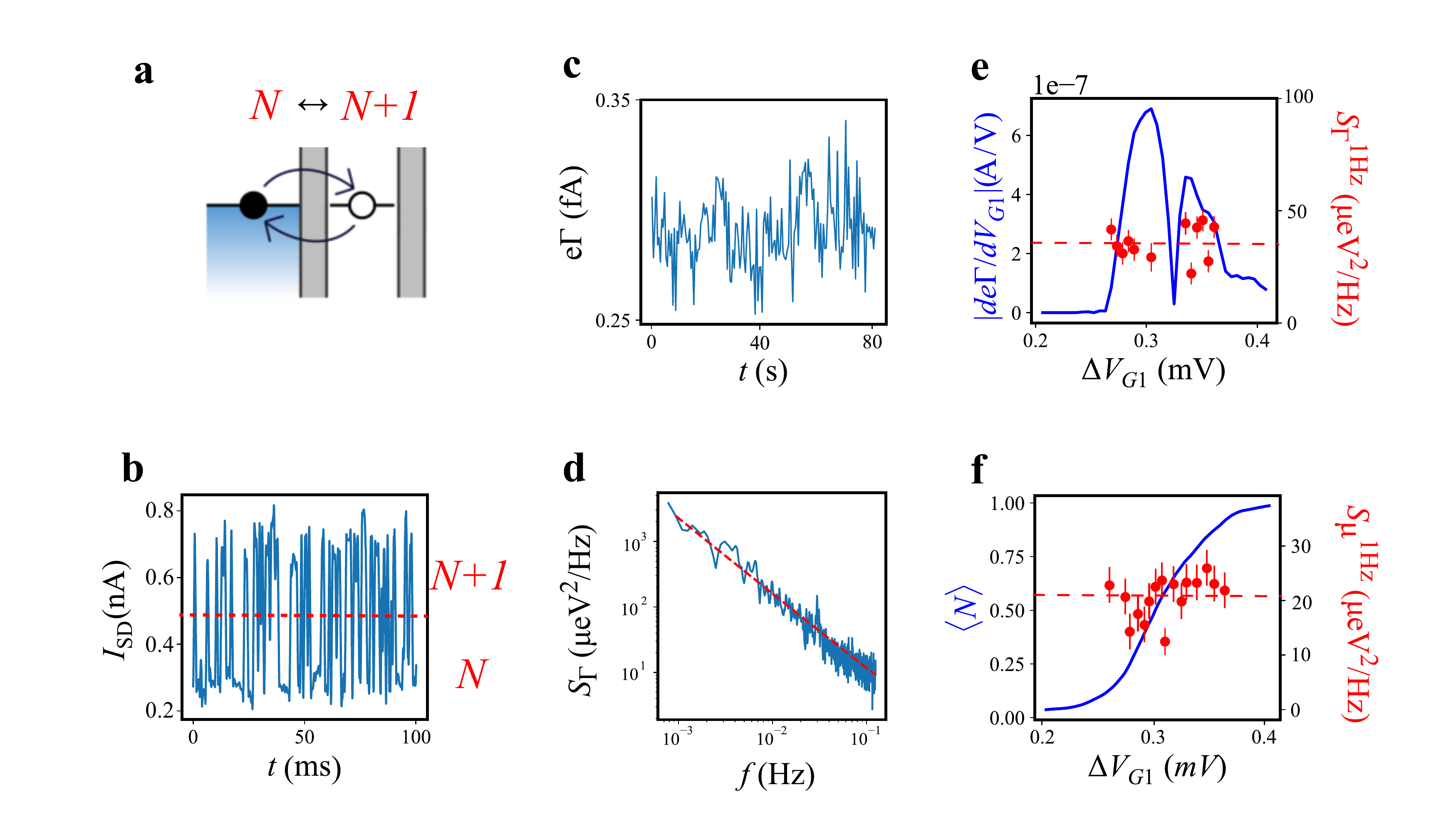}%
  \caption{
Measurement method for charge noise in a few-electron quantum dot. 
(a) Schematic of the quantum dot configuration to be probed. 
The first available energy level in the quantum dot is brought into resonance with the reservoir potential. 
The occupancy of the dot varies between N = 0 and N = 1.
(b) Current through the sensor dot measured over \SI{100}{ms}. 
A threshold is set to distinguish the two charge states of the probed dot 
(c) The variation in the equivalent current $e\Gamma$ is plotted over the course of a single measurement.
(d) Power spectral density obtained by Fourier transform of the plot in (c). 
%(f) The tunnel coupling $t_c$ is plotted as a function of the gate voltage. 
(e) In blue, charge sensitivity of the transition. In red, variation of the renormalized PSD at \SI{1}{Hz} as a function of the gate voltage for the same transition.
(f) The blue curve represents the evolution of the dot occupancy as a function of gate voltage. The derivative of such curve reflects the sensitivity of the chemical potential to charge noise. The red dots correspond to charge noise acting purely on the chemical potential and are extracted from $\braket{N}
(t)$ time traces.
}
  \label{fig:tunneling}%	
  \end{figure*}
%%%%%%%%%%%%%%%%%%%%%%%%%%%%%%%%%%%%%%%%%%%%%%%%%%%%%%%%%%%%%%%%%%%%%%%%
\section{Charge noise at the single electron level}

The charge noise of quantum dots is generally measured in the many-electron regime due to the simplicity of the measurement method \cite{yoneda2018quantum}.
However, spin qubits typically operate with few electrons.
In this configuration, the environment of the device can be significantly different.
The charge noise magnitude measured in the many-electron regime is therefore of limited value in determining the charge noise a qubit would experience.
To illustrate the dramatic dependence on the number of localised electrons, we have observed a drastic reduction of charge noise when the number of electrons trapped in the QD is increased from 10 to 30 (see suppl. mat Fig. \ref{fig:temperature}(a)).
The mechanism behind the reduction of the effect of charge noise is expected to be a combination of electric field screening due to the additional charges within the quantum dot, and an increase in the self-capacitance of the quantum dot through the geometric deformation of expansion.

The evaluation of charge noise in the single electron regime is usually done by measuring the dephasing rate of the spin when limited by charge noise.
This method requires coherent manipulation and readout of spins which is complex to implement for large scale characterization, and cannot readily be achieved, for example, at probe station temperatures.
We propose here a simpler method which allows measurement of the charge noise at the single electron level.
The method consists of probing single electron tunneling in real time, and investigating fluctuations in the tunnel rate and average dot occupancy.
We start by setting the chemical potential of the single electron in resonance with the Fermi sea (see Fig. \ref{fig:tunneling}(a)) and digitizing tunneling events by charge sensing, Fig. \ref{fig:tunneling}(b).
We choose a regime of sensor operation where the step size between the two charge states is strong enough to be immune from charge noise effect on the sensor itself.
We can then observe real-time tunneling and turn it into a current-time equivalent plot, Fig. \ref{fig:tunneling}(c), where $\Gamma$ is the tunneling rate.
Fig. \ref{fig:tunneling} (d) shows the corresponding noise spectrum after a Fourier transformation of the time trace and renormalization by the differential conductance. 
This plot corresponds to the optimal configuration of back-gate and top-gate voltages ($V_{BG} \approx 10 V$, $V_{TG} \approx 0 V$), which can reduce the charge noise at 1 Hz to the range of $1~\si{\micro\electronvolt}^2$/Hz (extrapolated from the fit).
The differential conductance is obtained by measuring the tunneling rate while scanning the gate G1. 
It leads to an effective Coulomb peak.
The derivative of this yields the differential conductance (see the blue curve in Fig. \ref{fig:tunneling}(e)).
On the same figure, we have extracted the charge noise at 1 Hz across the Coulomb peak for a typical configuration ($V_{BG}, V_{TG} \approx 0 V$), which leads to an average value of $40~\si{\micro\electronvolt}^2$/Hz, an order of magnitude higher noise compared to the many-electron regime (Fig. \ref{fig:1/f} (b)). 
As in the many electron regime, the charge noise is extremely dependent on the biasing conditions.\\

So far, the extraction of charge noise in both the many and few electron did not disentangle the contributions of tunnel barrier fluctuations and chemical potential fluctuations on the charge noise.
We now try to extract the charge noise purely acting on the chemical potential by probing the mean occupancy of the quantum dot, $\braket{N}$.
%Real time tunneling measurements allow us to extract the tunnel coupling to the leads as a function of time for a fixed gate voltage using $\Gamma_c=t^{-1}_{in}+t^{-1}_{out}$, with $t_{in}$ and $t_{out}$ the characteristic time for the electron to enter and leave the dot respectively. 
  \begin{figure}%
    \includegraphics[scale=0.6]{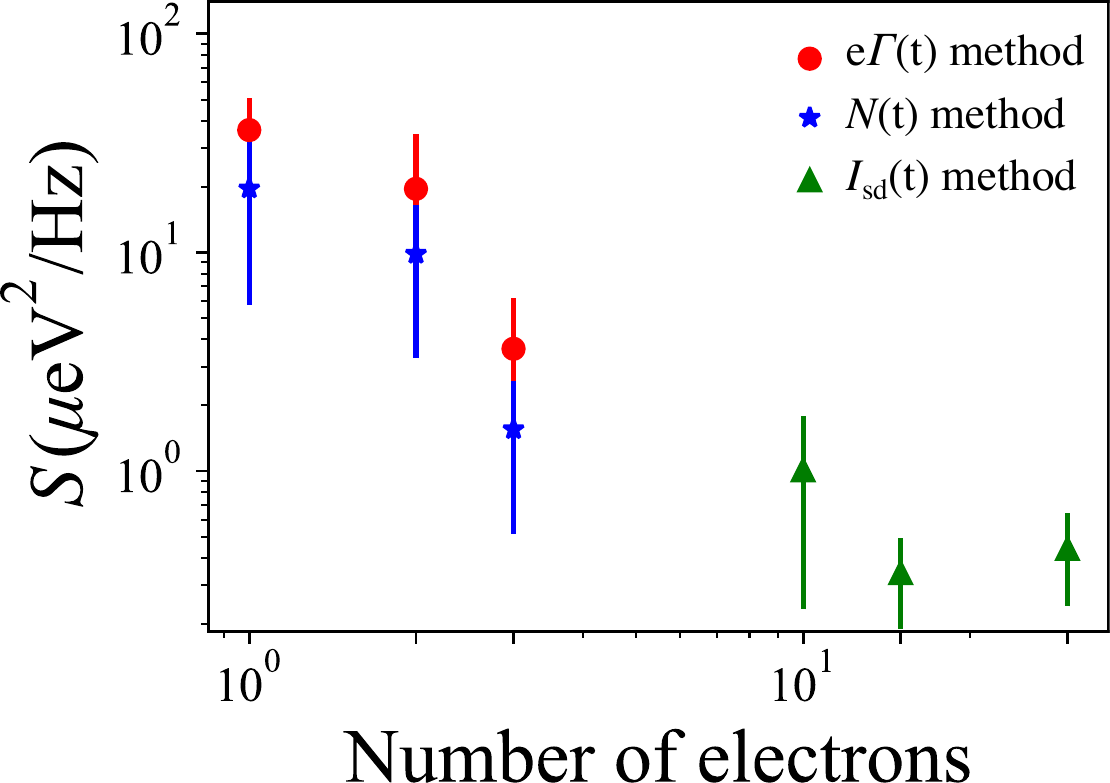}%
    \caption{
      Charge noise as a function of the dot occupancy. 
      The first three points for electron number 1, 2 and 3 are measured using the single shot method, deconstructed into the e$\Gamma (t)$ and $\langle N \rangle (t)$ contributions.
      The other points are measured in similar electrostatic configurations using a current measurement on a Coulomb peak. The values quoted here are for measurements conducted at a temperature of 360 mK.
  }
    \label{fig:ne}%	
    \end{figure}
    
$\braket{N}$ is extracted by averaging the real-time measurements using the expression $N = \frac{\tau_{N=1} + \tau_{N=0}}{\tau_{N=1}}$ with $\tau_{N=1}$ and $\tau_{N=0}$ the characteristic times during which the dot has one or zero electron.
To build the the power spectral density of the charge occupancy, the quantity $\braket{N}$ is measured as a function of time and Fourier-transformed. 
Further renormalization using $\frac{d\braket{N}}{dV_G}$ leads to the charge noise on chemical potential only, see \ref{fig:tunneling} (f). We obtain an average charge noise power of $20~\si{\micro\electronvolt}^2$/Hz.
This value is smaller than the one obtained from tunneling rate extraction.
This discrepancy is likely explained by accounting for charge noise in the tunnel barrier.
To extract this latter parameter, we would need a gate to control and extract the quantum dot sensitivity to tunnel coupling.
%Crucially, this value is non-negligible when compared to temperature on the time scale of a typical measurement.
%Thermal tunnelling is detrimental to spin readout and charge noise fluctuations on this order can have a similar effect, limiting or degrading the spin readout fidelity.
%Additionally, low-frequency fluctuations can cause the transition to shift implying the need to re-calibrate the readout during the measurement.

To further explore the dependence of the charge noise on the number of electrons, we measure similarly for the second and third electrons, which are presented in Fig. \ref{fig:ne} together with other points measured using the current fluctuation method.
We observe a clear decrease in the charge noise magnitude as the number of electrons increases in the quantum dot.
There are three different mechanisms to explain this behavior. 
The first one is the difference in electric field for the different electron numbers. 
For instance, measurements in the few electron regime require the presence of a charge detector and therefore different biasing conditions compared to the many electron regime.
The second potential explanation is a geometric consideration. 
As the number of electrons increases in the quantum dot, the volume of the quantum dot increases due to Coulomb interaction and modification of the confinement potential shape.
It leads to an increase in distance between the charge noise source, beyond the interface, and the average electron position, decreasing their dipolar interaction.
This assumes that the dominant fluctuators are located in the interface directly above the quantum dot.
The observed data is supported qualitatively by simulating the effect of a single TLS above the interface on the chemical potential of the quantum dot for three different quantum dot occupation (see supplementary material \ref{supinf:simu}).
Finally, screening effects can appear as the electron number increases leading to a reduction of the potential fluctuations, but this is expected to be a second order correction \cite{barnes2011}.

%For a single electron, the charge fluctuators induce electric field fluctuations and push the electron away from the center of its potential, causing it to experience significant noise.
%But as the number of electrons increases in the quantum dot, screening and geometric effects reduce the impact of the impurity fluctuation on the chemical potential \cite{barnes2011}, and the corresponding noise is reduced in magnitude.
%While a geometric model accounting for the decrease of charge noise with electron number failed to reproduce the experimental data, as they were taken in different electrostatic configurations, these results add to the potential advantages of operating the qubit device with higher numbers of electrons \cite{leon2020}.

%%%%%%%%%%%%%%%%%%%%%%%%%%%%%%%%%%%%%%%%%%%%%%%%%%%%%%%%%%%%%%%%%%%%%%%%
\section{Conclusion}
In conclusion, we have shown that the charge noise obtained from current spectroscopy gives valuable information on the effect of TLS on the QD chemical potential.
In particular, the charge noise strongly depends on the configuration of the quantum dot which can be tuned using the different gate potentials.
It also shows that the interfaces of the reservoirs are an important source of noise, which could be mitigated in future quantum technology applications via more precise control of the reservoir tunnel barriers.

Finally, we have shown that characterizing charge noise in the many electron regime is not sufficient to capture the full noise picture.
We have used a method based on single shot readout to extract the charge noise down to the single electron level.
It has allowed us to disentangle the pure chemical potential noise from the tunnel coupling fluctuations
Moreover, the method is experimentally simple to implement, requiring only charge detection of time-resolved tunneling at the single electron level.
It has allowed us to extract the charge noise susceptibility of the QD as a function of electron number which shows a decrease of charge noise at higher occupancy.
This last result adds to the potential benefits of operating a QD in the few electron regime to improve charge noise limited quantum dot based spin qubits \cite{leon2020}.
\section{Acknowledgment}
We acknowledge technical support from P. Perrier, H. Rodenas, E. Eyraud, D. Lepoittevin, I. Pheng, T. Crozes, L. Del Rey, D. Dufeu, J. Jarreau, J. Minet and C. Guttin. 
We thank Jing Li for fruitful discussions on the 1/f noise model.
C.S., B.K, and D.J.N. acknowledge the GreQuE doctoral programs (grant agreement No.754303). The device fabrication is funded through the Mosquito project (Grant agreement No.688539).
This work is supported by the Agence Nationale de la Recherche through the CRYMCO project and the CMOSQSPIN project. This project receives as well funding from the project QuCube(Grant agreement No.810504) and the project QLSI (Grant agreement No.951852)

%%%%%%%%%%%%%%%%%%%%%%%%%%%%%
\newpage 
\clearpage
\section{Appendix I: Simulation of charge noise for three different electron occupancy as a function of back gate}
\label{supinf:simu}

\begin{figure}[H]
    \includegraphics[width=\columnwidth]{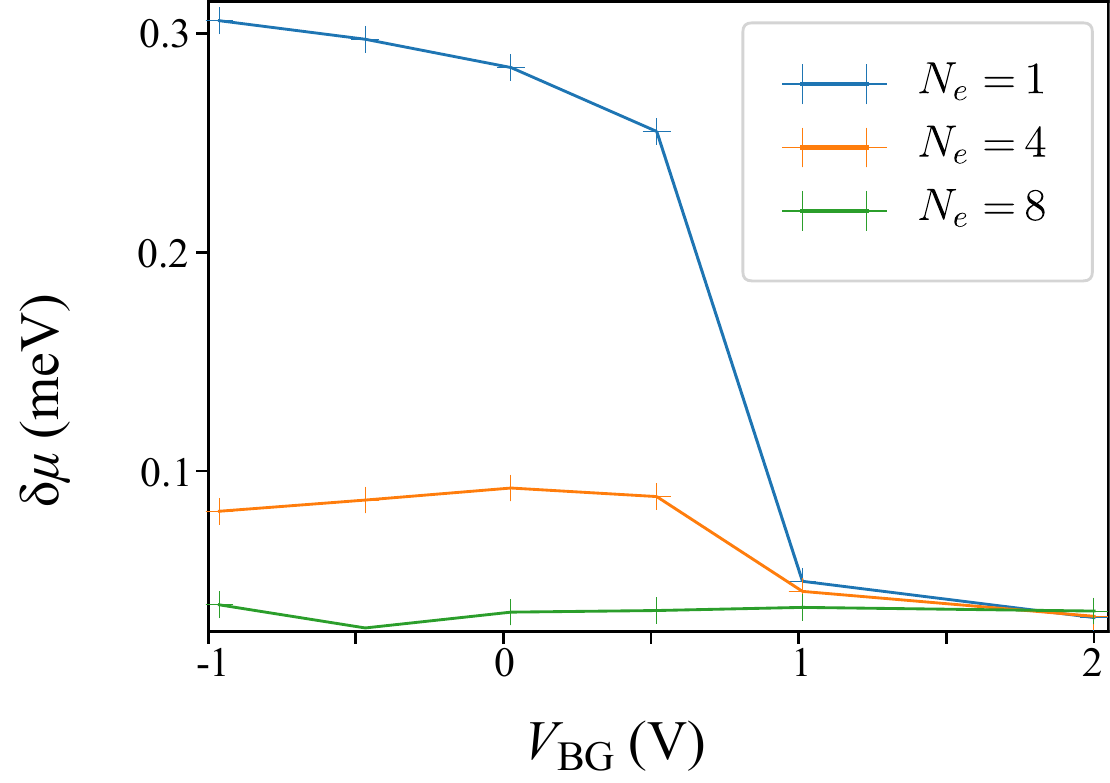}%
    \caption{
      Simulation of the variation of chemical potential induced by atwo level fluctuator as a function of the number of electrons in the dot and the backgate bias. It shows a qualitative agreement with the experimental variations of charge noise with both number of electrons and backgate voltage.
  }
    \label{fig:simu}%	
    \end{figure}

We compute the energies of $N$-electron configurations in QD1 in a self-consistent Schr\"odinger-Poisson approximation in which each electron moves in the potential created by the gates and by the mean-field density $(N-1)/N\times\rho(\mathbf{r})$, where $\rho(\mathbf{r})$ is the total electronic density and the factor $(N-1)/N$ accounts for the fact that each electron only interacts with $N-1$ others in the dot. Screening in the source and drain is accounted for in a linearized Thomas-Fermi approximation. To model the effect of a single fluctuator close to the dot, we compute the change of chemical potential when a single charge at the top interface of the channel moves by 1 nm perpendicular to the channel axis (see Fig. \ref{fig:simu}). The chemical potential fluctuations decrease when the back gate voltage is increased and the electron wave function moves from the top to the bottom interface, and when the number of electrons in the dot decrease, due to swelling of the dot (that extends further away from the charge) and to screening.

%%%%%%%%%%%%%%%%%%%%%%%%%%%%%%%%%%%%%%%%%%%%%%%%%%%%%%%%%%%%%%%%%%%%%%%%
\section{Appendix II: Assessing the effect of very low frequency charge fluctuators}
\begin{figure}%
\includegraphics[scale=0.4]{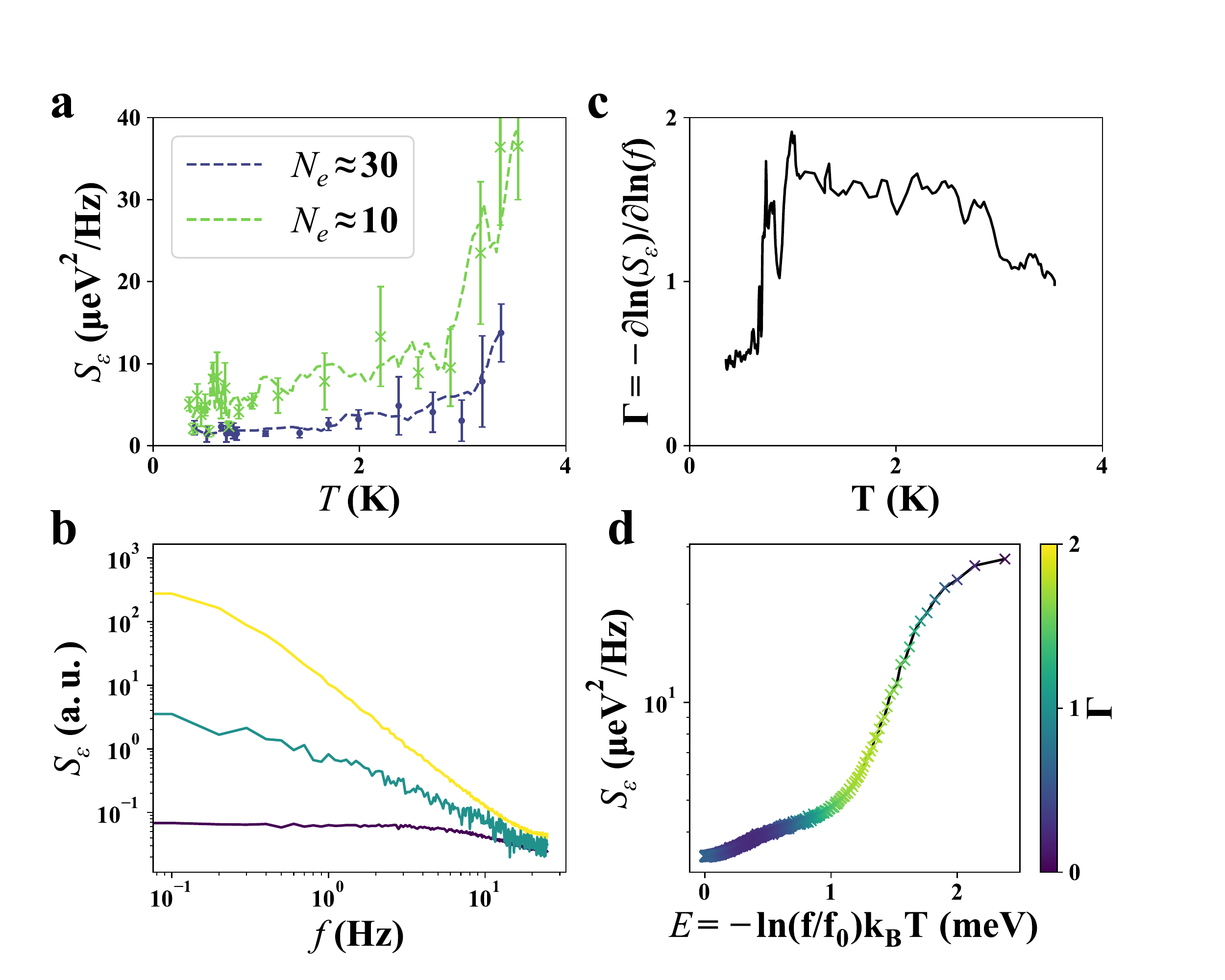}%
\caption{
(a) Power spectral density measured at $\SI{1}{Hz}$ as a function of temperature for two different dot configurations: the first visible coulomb peak (a few-electron configuration, with $N_e \approx 10$) (green), and the many-electron regime ($N_e \approx 30$) (purple).
$V_{TG}$ and $V_{BG}$ are kept constant.
The number of electrons is varied by increasing the on-site potential via plunger gate $V_{G1}$.
(b) Simulations of three different fluctuator configurations.
Low-frequency dominant fluctuators are simulated at $\SI{0.1}{Hz}$ (yellow), indicating a steep $1/f^2$ shape at $\SI{1}{Hz}$.
A distribution of many fluctuators at different frequencies (linear interpolation from $\SI{0.1}{Hz}$ to $\SI{100}{Hz}$) is simulated with equal amplitudes (green), demonstrating archetypal $1/f$ noise.
High-frequency dominant fluctuators are simulated at $\SI{10}{Hz}$ (purple), demonstrating a shifted lorentzian shape that tends towards $\partial \text{ln}(S_{\epsilon})/\partial \text{ln}(f) = 0$ at $\SI{1}{Hz}$.
(c) The measured $\partial \text{ln}(S_{\epsilon})/\partial \text{ln}(f)$ at $\SI{1}{Hz}$ as a function of temperature for the configuration $N_e \approx 10$.
(d) Simulation of the power density spectrum produced by two bistable fluctuators that satisfy the measured $\partial \text{ln}(S_{\epsilon})/\partial \text{ln}(f)$ from (c).
The simulated fluctuators have energies $\SI{0.1}{meV}$ and $\SI{1.5}{meV}$.
This curve gives the charge noise for any combination of $f$ and $T$ within the range of energy probed for this particular configuration.
The colour scale indicates the expected gradient of the related charge noise with frequency.
}
\label{fig:temperature}%	
\end{figure}
\subsection{Temperature dependence of charge noise}
As the characteristic fluctuation frequency of TLS follows an Arrhenius law $e^{-E_{\alpha} /k_B T}$, temperature can provide important information on the charge noise source structure \cite{ahn2021microscopic}.
Therefore, we investigate the temperature dependence of charge noise at \SI{1}{Hz} in the device.
For this purpose, we sweep the temperature from $\SI{400}{mK}$ to $\SI{4}{K}$.
In addition, as a control experiment we measure the charge noise at two different chemical potential configurations, corresponding to 10 and 30 electrons respectively.
Fig. \ref{fig:temperature}(a) presents the charge noise as a function of temperature for these two cases.
It shows a strong deviation from the expected linear dependence of charge noise \cite{petit2018spin}.
This is a signature of a non uniform distribution of the two-level system (TLS) energies.
To further explore the noise structure, we investigate the gradient of the power spectrum at $\SI{1}{Hz}$.
%%%%%%%%%%%%%%%%%%%%%%%%%%%%%%%%%%%%%%%%%%%%%%%%%%%%%%%%%%%%%%%%%%%%%%%%
\subsection{Reconstruction of TLS spectrum}

Fig. \ref{fig:temperature}(b) presents the simulated characteristic spectra of a group of TLSs evenly coupled to to the QD at a fixed temperature.
The green curve corresponds to the archetypal $1/f$ spectrum, which is the spectrum obtained by measuring a set of TLSs with a uniform distribution of characteristic frequencies.
The purple curve instead shows the power spectrum for a group of TLS with a distribution of characteristic frequencies centered at relatively high frequency.
The main consequence is that the gradient of the power spectrum is close to zero at $\SI{1}{Hz}$.
The last case (yellow curve) corresponds to a group of TLS with low characteristic frequency.
The $1/f^2$ dependence at $\SI{1}{Hz}$ orginates from the Lorentzian contribution of TLS to the power spectrum.
A real spectrum consists of a combination of these signatures dependent on the distribution of TLSs in the system.
Analysis of the gradient of the power spectrum at different frequencies can therefore give information about the fluctuators present in the system.
In order to probe a wide range of frequencies, we vary the temperature of the system.
We assume the activation energy $E_{\alpha}$ of a individual fluctuators to be fixed, and their characteristic frequency to be given by a kinetic process following an Arrhenius law $e^{-E_{\alpha} /k_B T}$ (see supplementary materials).

To probe the distribution of activation energies of the dominant fluctuators, we systematically extract the parameter $\Gamma = \partial \text{ln} S_{\epsilon} / \partial \text{ln}f$, the gradient of the power spectrum at $\SI{1}{Hz}$ in logarithmic scale.
Fig. \ref{fig:temperature}(c) shows the evolution of $\Gamma$ as a function of temperature.
From this, we can qualitatively reconstruct the full spectrum of TLSs at a given temperature based on the extracted $\Gamma$.
Fig. \ref{fig:temperature}(d) is a simulation of the charge noise using two groups with activation energy distributions centered at $\SI{0.1}{meV}$ and $\SI{1.5}{meV}$.
One group has a high characteristic frequency, which likely corresponds to the dopants on the edge of the reservoir that dominate the increase in charge noise with top gate.
The other has lower characteristic frequency, which is expected to be seen with TLSs located at the interfaces or embedded in the gate stack or buried oxide.
\subsection{Arrhenius activation energy for identified TLS}

  \begin{figure}%
    \includegraphics[width=\columnwidth]{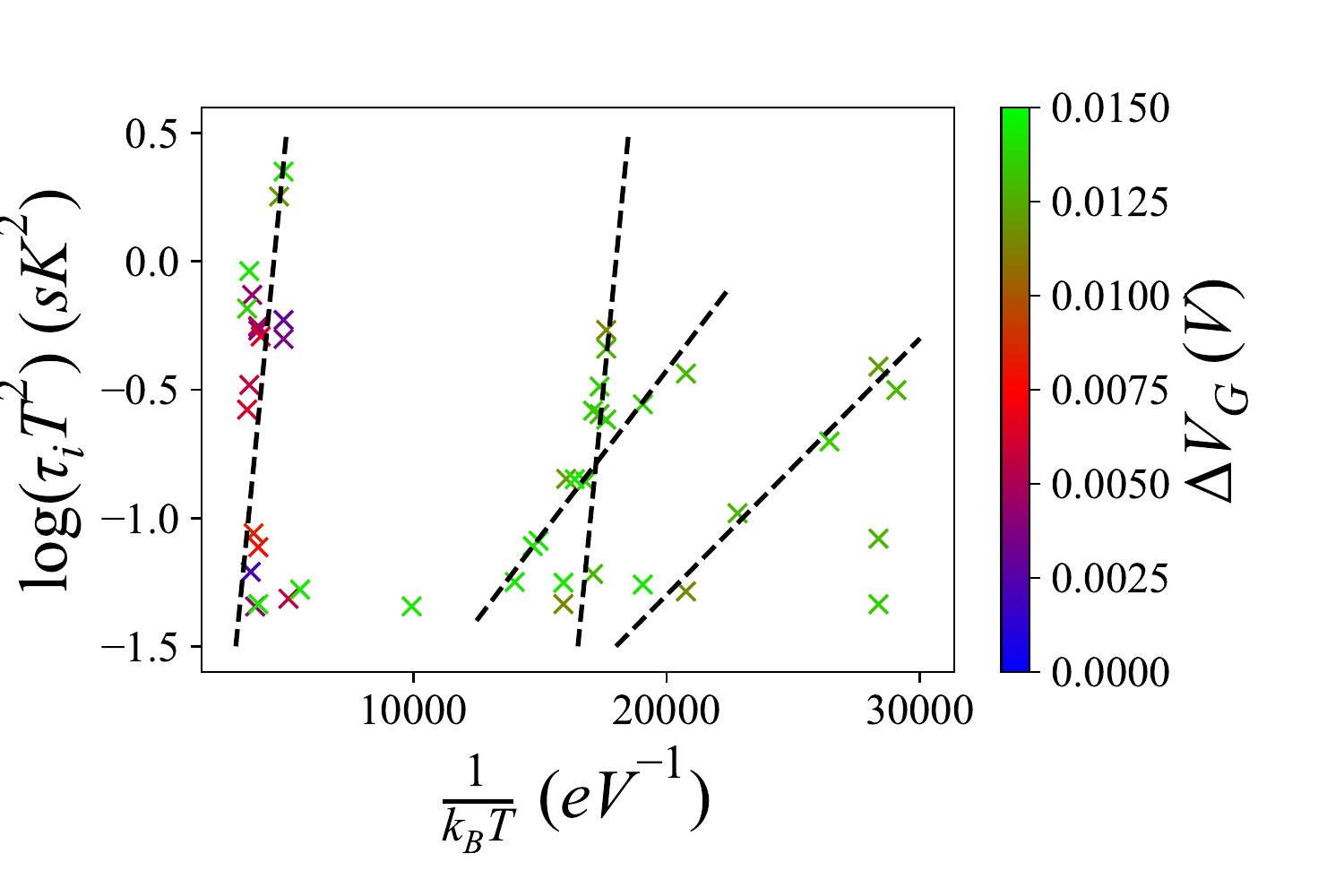}%
    \caption{
      Arrhenius diagram of the frequency of individual fluctuators detected in a series of charge noise measurements made over a temperature range from \SI{0.3}{K} to \SI{4}{K}.
      The colour scale gives the voltage offset from a reference point at the coulomb peak for each detected dominating fluctuator.
      Points which are caused by the same fluctuator are expected to have approximately the same voltage offset.
  }
    \label{fig:arrhenius}%	
    \end{figure}

An Arrhenius plot was used to confirm the individual fluctuator energies extracted using the energetic simulation, and to obtain their energies independent of the characteristic frequency. 
For a single quantum dot configuration, charge noise measurements were taken over a temperature range from \SI{0.3}{K} to \SI{4}{K}.
The time constant of these dominant fluctuators was determined by fitting the noise spectrum with a gaussian expression and extracting the central frequency.
The time constant was then plotted against the temperature (Fig. \ref{fig:arrhenius}).
Two groups of fluctuators were observed to change linearly with temperature.
The energy of these fluctuators was determined by their gradient, giving \SI{0.15}{meV} and \SI{1.35}{meV} respectively, in good agreement with the values obtained through simulation.

\end{document}